\documentclass[proceedings]{JHEP} 

\usepackage{epsfig}			

\newbox\mybox
\newcommand\fverb{\setbox\mybox=\hbox\bgroup\verb}
\newcommand\fverbdo{\egroup\medskip\noindent\fbox{\unhbox\mybox}\ }
\newcommand\fverbit{\egroup\item[\fbox{\unhbox\mybox}]}

\font\beeg=cmr17 scaled 1600		
\newcommand\init[1]{\setbox\mybox=\hbox{{\beeg #1}~}%
		   \noindent\global\hangindent=\wd\mybox\global\hangafter-2%
		   \sc\smash{\llap {\lower 13.2pt \box\mybox}}}

\title{Semileptonic B Decays from CLEO}

\author{Karl M. Ecklund\thanks{Representing the CLEO Collaboration}\\
	Newman Laboratory of Nuclear Studies\\
        Cornell University, Ithaca, New York, USA\\
	E-mail: \email{kme@mail.lns.cornell.edu}}

\conference{Heavy Flavours 8, Southampton, UK, 1999}

\abstract{I report results in semileptonic decays of $B$ mesons from
the CLEO collaboration, with a focus on recent results.  Results for
exclusive reconstruction of $B\to D^*\ell\nu$, $B\to D\ell\nu$ and
$B\to\rho\ell\nu$ are given including the $q^2$ dependence of the form
factors.  These results are used to measure $|V_{cb}|$ and $|V_{ub}|$.
Two preliminary analyses using inclusive techniques measure the lepton
momentum spectrum and hadronic recoil mass spectrum in $B\to
X_c\ell\nu$ decays, showing promise for future precision measurements
of $|V_{cb}|$.  }

\begin{document} 

\maketitle 


Study of semileptonic decays of $B$ mesons allows measurement
of the CKM matrix elements $|V_{ub}|$ and $|V_{cb}|$
\cite{kme:Ckm,kme:cKM}.  Accurate measurement of CKM matrix elements
becomes increasingly important as we enter the era of the
$B$-Factories and studies of $CP$ violation in $B$ meson decays.

In the Standard Model, $CP$ violation comes about through a non-zero
phase in the CKM matrix, which describes quark mixing in weak decays.
Decays of the $b$-quark in particular will be key to our understanding
of $CP$ violation and flavour physics.  In the well known unitarity
triangle, $|V_{cb}|$ sets the overall scale for the length of the
sides, and $|V_{ub}|$ determines the length of one side.  Precise
determinations of both will be needed to complement the measurement of
the angles of the unitarity triangle in view at the $B$-Factories.
The goal is to over-constrain the unitarity triangle and test the
Standard Model.

\FIGURE{
\epsfig{file=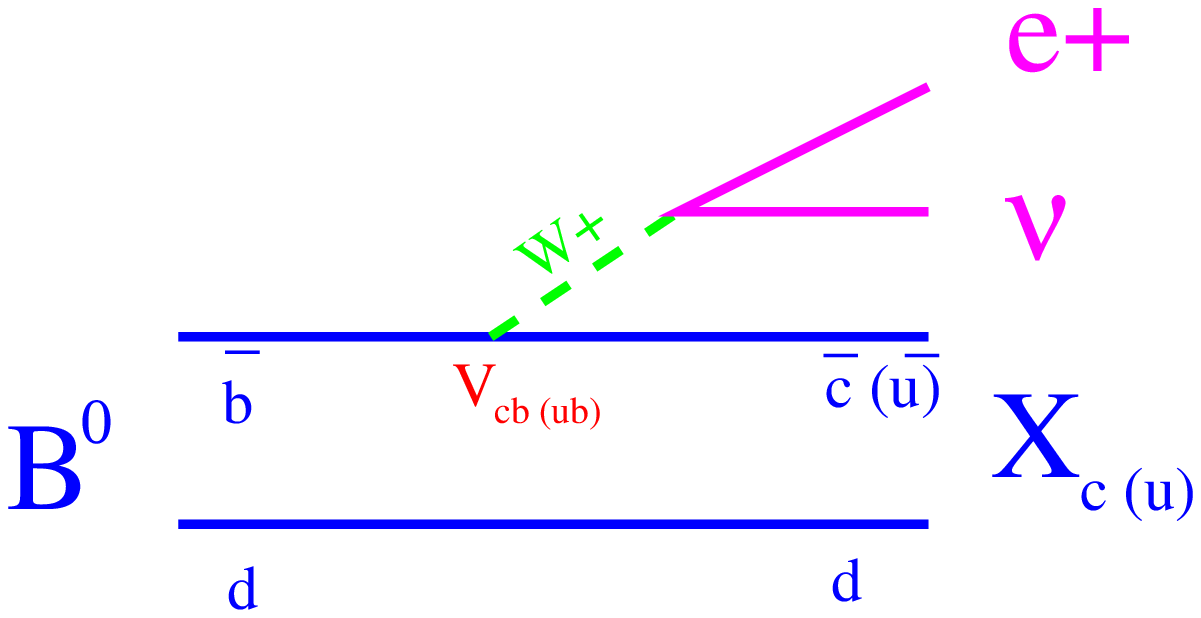,width=6.8cm}
\caption{Diagram for semileptonic $B$ decay.}
\label{f:kme:semileptonic}
}
In principle, CKM matrix elements can be studied in any weak decay
mediated by the $W$ boson.  Semileptonic decays (see
figure~\ref{f:kme:semileptonic}) offer the advantage that
the leptonic current is calculable and QCD complications only arise in
the hadronic current.  Unlike hadronic decays, there are no final state
interactions and only one amplitude contributes to the decay.  One
still needs some understanding of the strong interaction.  Heavy Quark
Effective Theory and other approaches to QCD calculations such as light
cone sum rules and lattice QCD offer detailed and robust predictions
for the QCD dynamics in heavy quark decay.  These predictions allow
measurement of $|V_{ub}|$ and $|V_{cb}|$ with reasonable precision.
Data from experiment are beginning to test our understanding of heavy
quark decay.  As theory and experiment interact, an improved
understanding will lead to more precise measurements.

In what follows, I present results in exclusive and inclusive
semileptonic decays from the CLEO collaboration.  We study the
exclusive decays $B\to D^{(*)}\ell\nu$ to measure $|V_{cb}|$ and the
form factors which describe the role of the strong interaction in the
decay.  In $B\to \pi(\rho)\ell\nu$ we study the form factor for $B$
transitions to light mesons and measure $|V_{ub}|$.  Inclusive
measurements of $B\to X_c\ell\nu$ offer comparable precision on
$|V_{cb}|$ with new techniques that constrain phenomenological
parameters using decay spectra.

\section{CLEO Experiment}
CLEO is a $4\pi$ solenoidal detector located at the interaction region
of the Cornell Electron Storage Ring (CESR).  CESR is a symmetric $e^+e^-$
collider operating on the $\Upsilon(4S)$ resonance at a center of
mass energy of 10.58 GeV, just above the threshold for $B \bar B$
production.  $\Upsilon(4S)$ decays are essentially 100\% $B_d^0 {\bar
B_d^0}$ and $B^+B^-$ pairs.  At threshold the $B$'s are
produced nearly at rest: $p_B\approx 300$ MeV/$c$.  In addition to
$\Upsilon(4S)$ production with a cross section of 1.0 nb, there is
continuum production (3.0 nb) of $e^+ e^- \to $ hadrons.  CLEO also
collects data 60 MeV below the $\Upsilon(4S)$ for use in subtraction
of this continuum from on-resonance data.

The central region of the CLEO detector consists of three concentric
cylindrical drift chambers, a scintillator time-of-flight system and a
CsI calorimeter all inside a superconducting coil and 1.5 T magnetic
field.  Endcap time-of-flight and CsI calorimeters provide forward and
backward coverage for a total of 95\% of the solid angle.  The drift
chambers provide excellent tracking and momentum resolution, and the
calorimeter has excellent photon and $\pi^0$ identification.  In the
flux return for the superconducting solenoid, proportional tube
counters provide muon identification at depths of 3, 5 and 7
interaction lengths.  The CLEO detector is described in detail
elsewhere~\cite{kme:nim}.

\section{$B\to D^*\ell\nu$}
As an example of how one can extract $|V_{cb}|$ from an exclusive
semileptonic $B$ decay, consider $B\to D^*\ell\nu$.  The differential
decay rate is given by equation \ref{e:kme:dgdw}.
\begin{eqnarray}
{d\Gamma\over dw}  = { G_F^2 \over 48 \pi^3} |V_{cb}|^2 
                     {\cal F}_{D^*}^2(w) {\cal G}(w)
\label{e:kme:dgdw}
\end{eqnarray}
Here $w = (M_B^2 + M_{D*}^2 - q^2)/(2M_B M_{D*})$ is the inner product
of the heavy quark four velocities, which is linearly related to
$q^2$, the invariant mass of the $W$.  The function
\begin{eqnarray}
{\cal G}(w)&=&M_{D^*}^3 (M_B - M_{D^*})^2 \sqrt{w^2-1} (w+1)^2 \nonumber
\\ 
&\times& \left[ 1+ {4w\over w+1} {1-2wr + r^2 \over (1-r)^2}\right]
\label{e:kme:dgdw_g}
\end{eqnarray}
with $r=M_{D^*}/M_B$, is a function of $w$ and the relevant masses.  The
function ${\cal F}_{D^*}(w)$ is the form factor for the $B$ to $D^*$
transition.  

In the limit of very heavy quarks ($M_{b,c}\to\infty$), ${\cal F}(w)$
becomes the universal Isgur-Wise function, and ${\cal F}(1)=1$.  That
is, the form factor is known with absolute normalization at the point
of zero recoil, or $w=1$.  For finite heavy quark masses the
corrections may be computed in the framework of Heavy Quark Effective
Theory (HQET).  The QCD corrections are computed in a perturbation
theory using $\Lambda_{\rm QCD}/M_b$ and $\alpha_s$ as expansion
parameters.  For $B\to D^*\ell\nu$ the first order correction vanishes
exactly (known as Luke's theorem \cite{kme:luke}), 
and the coefficients have been
computed to order $\alpha_s^2$ and $1/M_b^2$
\cite{kme:f1neu,kme:f1falk,kme:f1shif,kme:f1czar}:
\begin{eqnarray}
{\cal F}_{D^*}(w=1)=0.91 \pm 0.03. \nonumber
\end{eqnarray}
The value above is in agreement with the assessment of {\em The BaBar
Physics Book} authors \cite{kme:babarbook}, but Bigi has recently
given a central value of 0.89, with substantially larger
uncertainty~\cite{kme:f1bigi}.

In an experiment one measures the decay rate as a function of $w$
and extrapolates to $w=1$.  At this kinematic point, the $D^*$ has no
momentum in the rest frame of the $B$ meson.  At CLEO, where the $B$'s
are produced near threshold, the momentum of the resulting slow pion
from $D^{*+}\to D^0\pi^+$ is very small.  The efficiency for
reconstructing the slow pion approaches zero as one approaches the
zero-recoil point, making the extrapolation more difficult.  This
experimental difficulty is avoided for $B^-\to D^{*0}\ell\bar\nu$,
where a slow $\pi^0$ may be reconstructed even at very small momenta.

As the kinematically allowed range of $w$ is small ($w\in [1.0,1.5]$),
the form factor is approximated as a Taylor expansion about $w=1$.
\begin{equation}
{\cal F}(w) = {\cal F}(1) (1 + (w-1)\rho^2 + c(w-1)^2)
\end{equation}
CLEO has measured the $B\to D^*\ell\nu$ decay rate as a function of
$w$ as shown in figure \ref{f:kme:dslnu_w} \cite{kme:Dslnu95}.
\FIGURE{
\epsfig{file=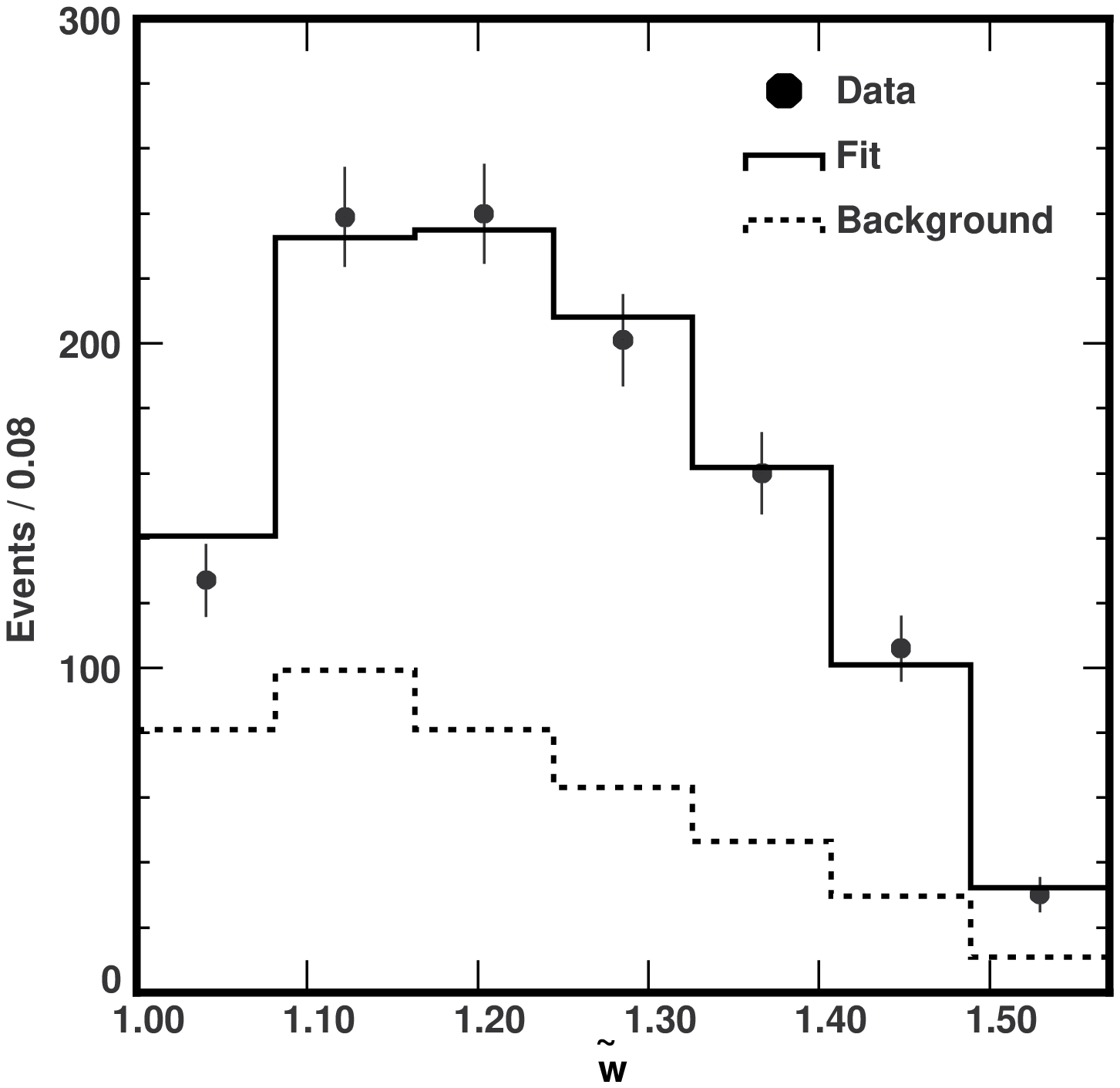,width=6.8cm}
\epsfig{file=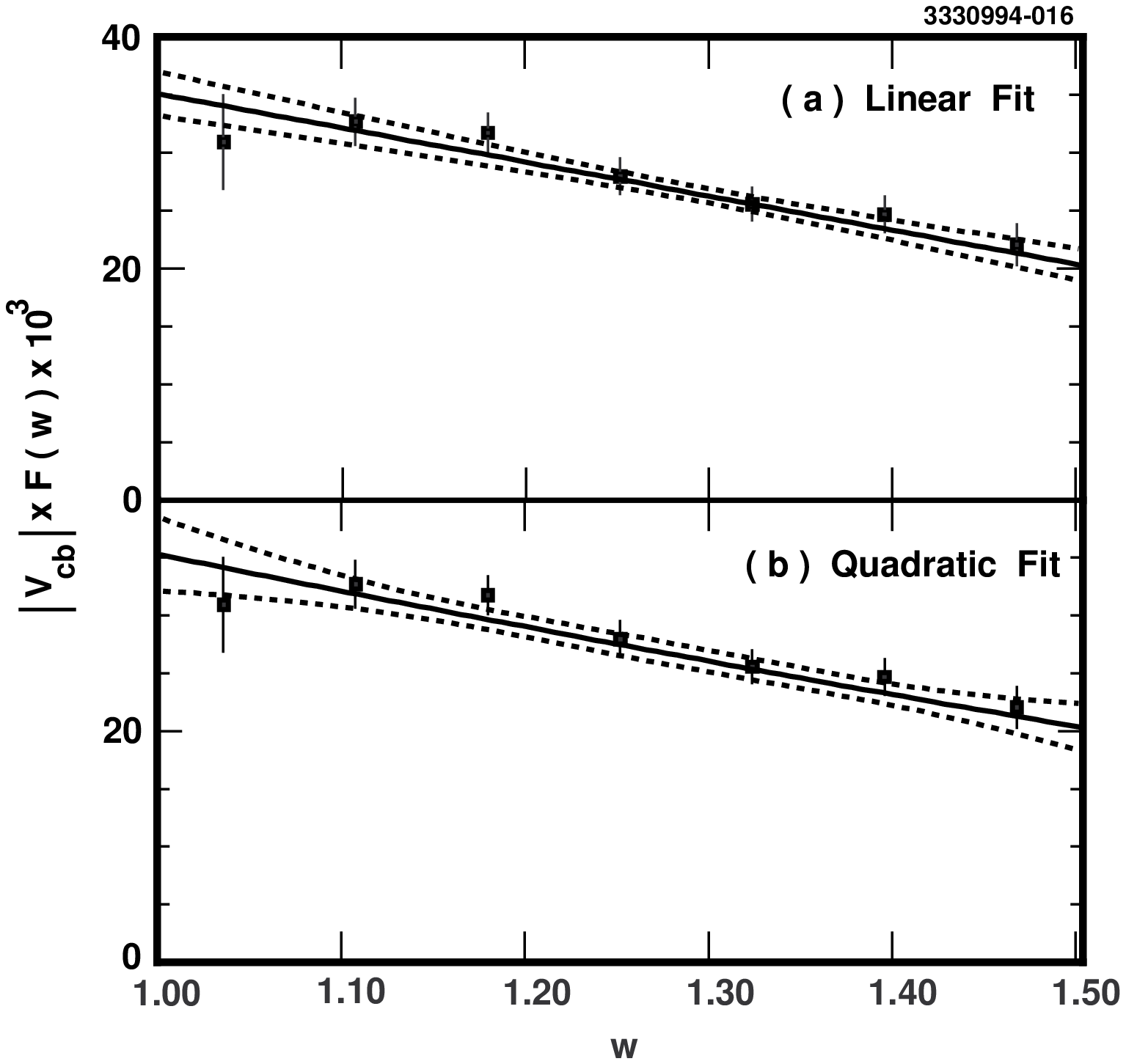,width=6.8cm}
\caption{The left side shows the differential yield $dN/dw$ for
$B^-\to D^{*0}\ell^-{\bar\nu}$ and ${\bar B^0} \to
D^{*+}\ell^-{\bar\nu}$.  The fit shown is to a linear form factor.
The background is mostly due to combinatoric $D^*$ candidates.  
On the right ${\cal F}(w)|V_{cb}|$ is plotted vs $w$. Data points
are overlayed with fit results (solid line) for linear (a) and
quadratic (b) form factors.  The dashed lines show the contours for 1
$\sigma$ variations of the fit parameters.}
\label{f:kme:dslnu_w}
}
The figure also shows ${\cal F}(w) |V_{cb}|$ versus $w$
(the kinematic factors have been divided out) showing that experiment
currently does not differentiate between a linear and quadratic
dependence on $w$ for the form factor.
Taking the linear fit, one may read off the value for 
${\cal F}(1)|V_{cb}|$ at the intercept, 
\begin{eqnarray}
{\cal F}_{D^*}(1) |V_{cb}| = 0.0350 \pm 0.0019 \pm 0.0018 \pm
0.0003. \nonumber
\end{eqnarray}
The uncertainties are statistical, the systematic (dominated by slow pion
efficiency) and due to the $B$ lifetime.
I have updated the result using the PDG98 $B$ lifetimes~\cite{kme:PDG98}.
Using ${\cal F}(1)=0.91 \pm 0.03$, this gives
\begin{eqnarray}
|V_{cb}|=0.0385 \pm 0.0029 \pm 0.0013. \nonumber
\end{eqnarray}

This result is based on 1.6 million $B\bar B$ pairs.  CLEO currently
has nearly 10 million $B\bar B$ events, so substantial improvement in
this measurement may be expected.  LEP experiments also use this
technique to measure $|V_{cb}|$ \cite{kme:calvi} using a smaller
number of $B$ decays, with somewhat different experimental systematic
uncertainties.  For CLEO, the leading systematic comes from modelling
of the slow pion efficiency.  At LEP, contributions from semileptonic
decay to higher excited charm mesons ($B\to D^{**} \ell \nu$), which
are poorly known, lead to non-negligible systematic uncertainties.

\section{$B\to D\ell\nu$}
The decay $B\to D\ell\nu$ can be analyzed in the same way as the
$D^*\ell\nu$ decay described above, although the precision with which
$|V_{cb}|$ can be determined is not as good because of a smaller
branching fraction, larger backgrounds and an additional kinematic
suppression factor of $w^2-1$.
(Compare equations~\ref{e:kme:dgdw_g} and \ref{e:kme:dgdw_dlnu}.)
Nonetheless it provides complementary information and provides a test
of HQET predictions for relationships between the form factors for
semileptonic decays of pseudoscalar ($B$) to pseudoscalar ($D$)
and pseudoscalar to vector ($D^*$).  In HQET to first order there is a
common form factor, the Isgur-Wise function $\xi(w)$.

CLEO has two recent analyses of $B\to D\ell\nu$.  The first
\cite{kme:Dlnu97} uses a neutrino reconstruction technique to
completely reconstruct the decay chain.

\FIGURE{\epsfig{file=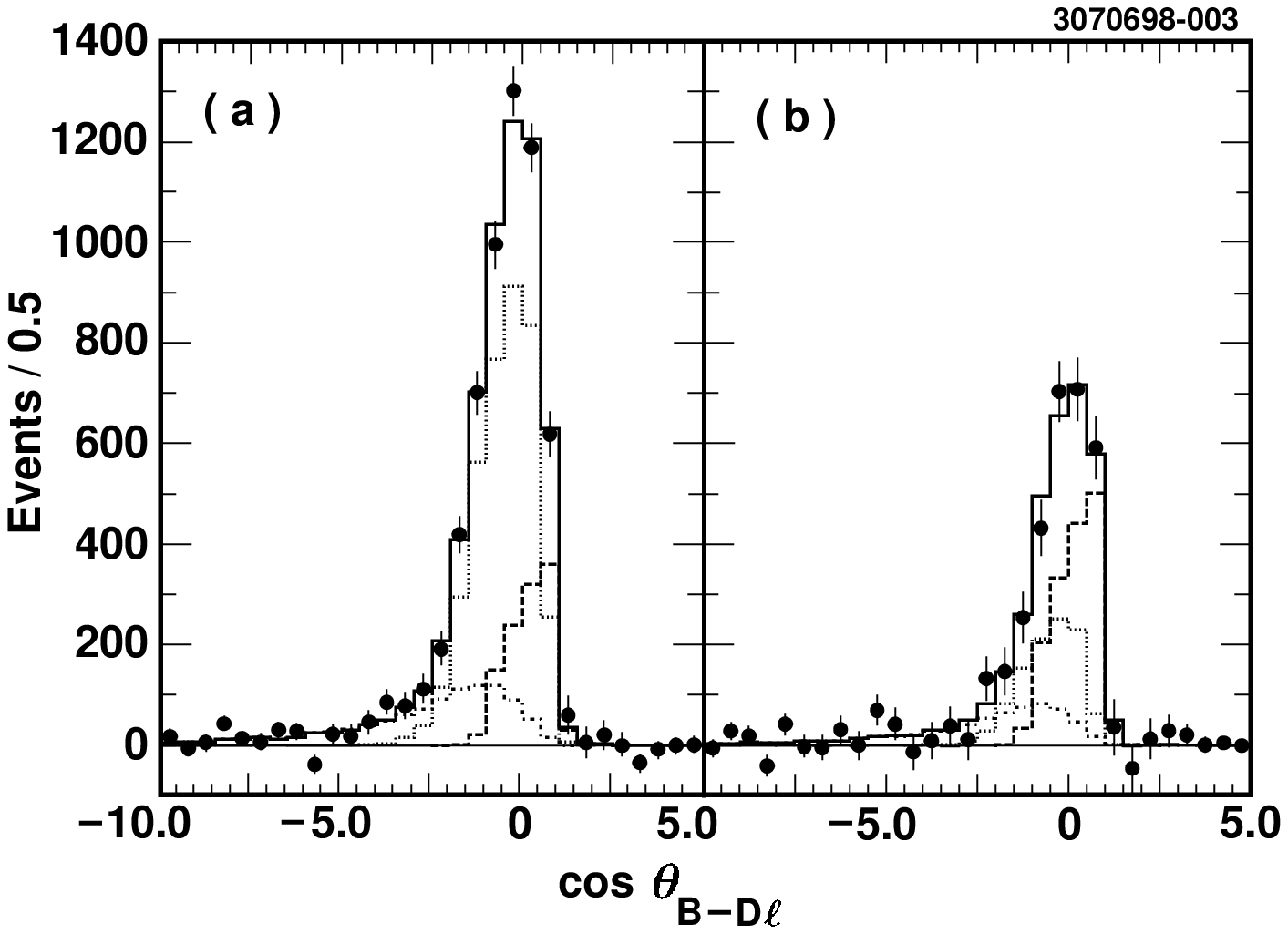,width=3.5in}
\caption{The $\cos\theta_{B-D\ell}$ distribution for (a) $D^0X\ell\nu$ and
(b) $D^+X\ell\nu$ candidates.  The data (solid circles) are overlayed
with simulated $B\to D\ell\nu$ decays (dashed histogram), $B\to
D^*\ell\nu$ decays (dotted histogram), $B\to
D^{**}\ell\nu+D^{(*)}\pi\ell\nu$ decays (dash-dotted histogram), and
their total (solid histogram).}
\label{f:kme:dlnu_cby}
}
In the second analysis \cite{kme:Dlnu99}, we identify events
containing a $D^+$ or $D^0$ (and charge conjugates) and an electron or
muon ($\ell$).  The $D$-$\ell$ combinations give a sample including
$B\to D\ell\nu$, $B\to D^*\ell\nu$, $B\to D^{**}\ell\nu$ and $B\to
D^{(*)}\pi\ell\nu$.  We separate $B\to D\ell\nu$ from the other
semileptonic modes using the energy and momentum of the particle(s)
recoiling against the $D$-$\ell$ pair.  The yield of $D\ell\nu$ events
in bins of $q^2$, the invariant mass of the virtual $W$, gives
information on the partial width and form factors in the decay $B\to
D\ell\nu$.

We reconstruct $D$ candidates in the modes $D^0\to K^- \pi^+$ or
$D^+\to K^-\pi^+\pi^+$, separating $K$ and $\pi$ tracks by using
time-of-flight and drift chamber $dE/dx$ measurements.  To suppress
$D$'s from the continuum, we require the $D$ candidates to have
momentum $p_D < 2.5$ GeV/$c$.  We select electron candidates of
momentum $0.8 < p_\ell < 2.4$ GeV/$c$ using the CsI calorimeter.
Muon candidates must have associated hits in the muon counters,
penetrating at least 5 interaction lengths of material, which
increases the lower momentum cut for muons to 1.4 GeV/$c$.  For 90\%
of signal $D\ell\nu$ events, the lepton and $D$ lie in opposite
hemispheres; we require this of all $D$-$\ell$ pairs.

For each $D$-$\ell$ pair we compute $\cos\theta_{B-D\ell}$, the angle
between the $D\ell$ momentum and the $B$ momentum assuming that the only
missing particle is a massless neutrino.
\begin{equation}
\cos\theta_{B-D\ell}={2 E_B E_{D\ell} - M_B^2 -M_{D\ell}^2 
                      \over 2 |{\bf p}_B| |{\bf p}_{D\ell}|}
\label{e:kme:cby}
\end{equation}
For $B\to D\ell\nu$ decays $\cos\theta_{B-D\ell}$ lies between 1 and
-1.  When final state particles other than the neutrino are missing,
it is shifted towards negative values.  Thus we may use this quantity
to distinguish $D\ell\nu$ from $DX\ell\nu$.  Before doing so other
backgrounds must be subtracted.

Background sources yielding a $D$-$\ell$ pair may arise from 
(1) random $K\pi(\pi)$ combinations (fake $D$), 
(2) a $D$ paired with a lepton from the other $B$ decay (uncorrelated), 
(3) a $D$ paired with a lepton that is a granddaughter of the same $B$
(correlated), 
(4) misidentification of a hadron as a lepton, or 
(5) $e^+e^-\to q\bar q$ events.  We remove backgrounds from fake $D$
candidates by using events in the $D$ mass sidebands.
The uncorrelated background contribution is estimated from our data by
flipping the direction of leptons in the same hemisphere as the $D$
candidate.  The small amount of correlated background ({\em e.g.} from
$B\to D^{(*)}\tau\nu$, $\tau\to\ell\nu\bar\nu$) is removed using Monte
Carlo (MC) simulation.  Fake leptons and continuum events are
subtracted using measured fake rates and off-resonance data.

\TABLE{
\caption{Summary of $B\to D\ell\nu$ form factor fits.}
\label{t:kme:dlnu_ff}
\begin{tabular}{lccc}
Form Factor   & $\rho_D^2$               & $c_D$
& $ {\cal F}_D(1)|V_{cb}|10^2$ \\ \hline
Linear        & $0.76\pm0.16\pm0.09$            
& - 
& $4.05\pm0.45\pm0.32$ \\

Free curvature& $0.77^{+1.18}_{-2.83}\pm0.09$
&$0.01^{+1.70}_{-3.96}\pm0.001$  
& $4.05^{+1.51}_{-1.63}\pm0.32$ \\

Boyd          & $1.30\pm0.27\pm0.16$            
& $1.21\pm0.31\pm0.15$ 
& $4.48 \pm 0.61\pm0.37$ \\

Caprini       & $1.27 \pm 0.25 \pm 0.15$ 
& $1.18\pm0.26\pm0.14$ 
& $4.44\pm0.58\pm0.36$ \\
\end{tabular}
}

In figure~\ref{f:kme:dlnu_cby} the resulting $\cos\theta_{B-D\ell}$
distributions are shown along with a fit to the data.  We model the
distributions in the fit using MC simulation and various models for
$b\to c$ semileptonic decay: for $B\to D\ell\nu$ we use ISGW2
\cite{kme:isgw1,kme:isgw2}; for $B\to D^*\ell\nu$ we use CLEO form
factors \cite{kme:Dslnu95,kme:cleoFF2}; for $B\to D^{**}\ell\nu$ we
use ISGW2; and for non-resonant $B\to D^{(*)}\pi\ell\nu$ we use the
results of Goity and Roberts \cite{kme:goity}.
To extract form factor results 
we perform the fit to $\cos\theta_{B-D\ell}$ in ten bins
of the HQET variable $w=(M_B^2+M_D^2-q^2)/(2M_B M_D)$, where $q^2$ is
the invariant mass of the $D$-$\ell$ pair.  The $D\ell\nu$ yield in
each $w$ bin is shown in figure~\ref{f:kme:dlnu_w}.  We fit the
differential decay rate
\begin{equation}
{d\Gamma\over dw}=
{G_F^2 |V_{cb}|^2 \over 48 \pi^3} (M_B +M_D)^2 M_D^3 (w^2-1)^{3/2}
{\cal F}_D^2(w)
\label{e:kme:dgdw_dlnu}
\end{equation}
assuming different form factors ${\cal F}_D(w)$.  The fit accounts for
detector acceptance and smearing in the reconstruction of $w$ due to
motion of the $B$ and detector resolution ($\sigma_w=0.026$).
\FIGURE{
\epsfig{file=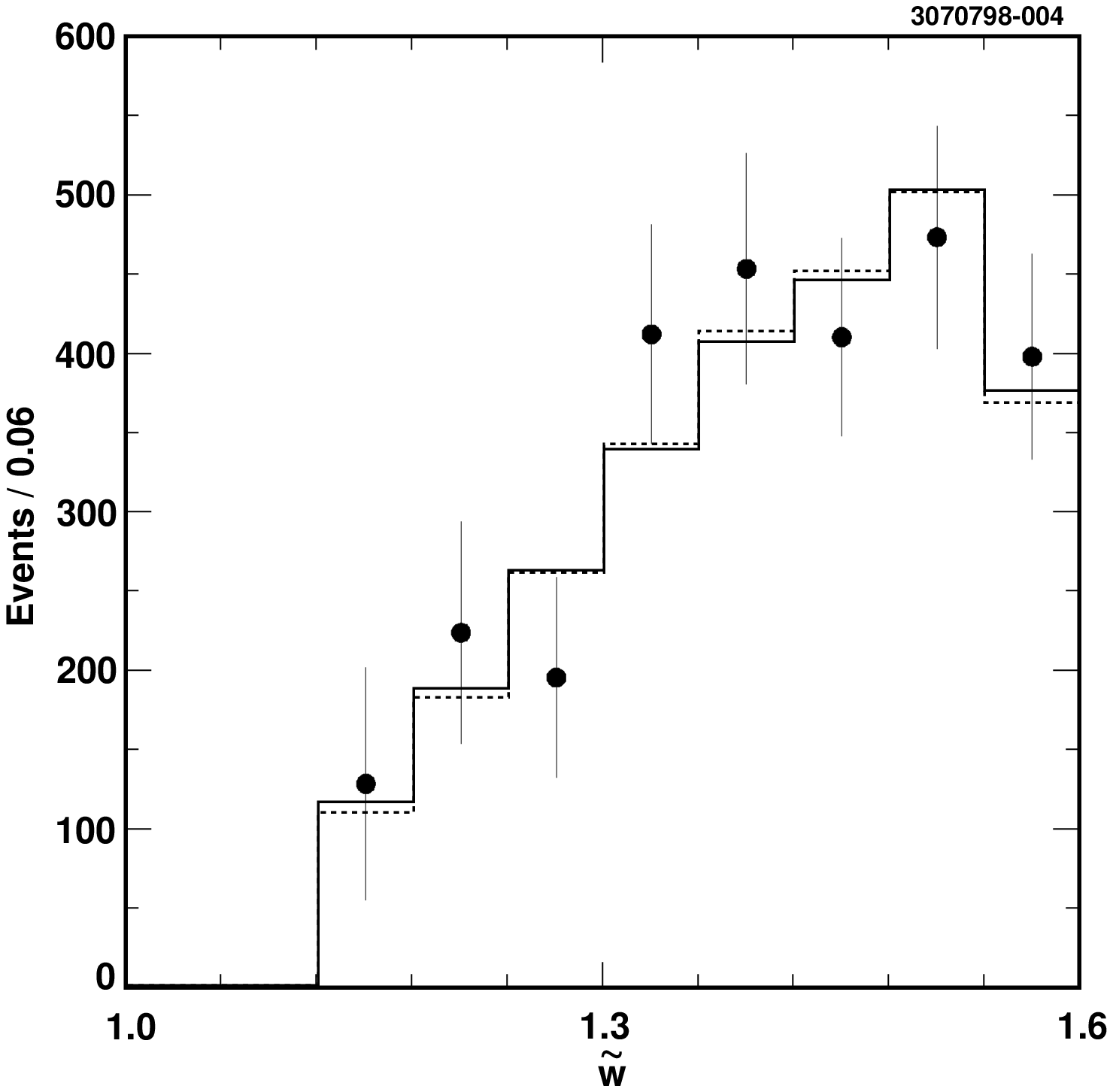,width=2.5in}
\caption{The sum of $B^-\to D^0\ell\bar\nu$ and $\bar{B^0}\to
D^+\ell\bar\nu$ yields as a function of $\tilde{w}$, for the data
(solid circles) and using the best fit linear form factor (dashed
histogram) or dispersion relation inspired form factor of Boyd {\em et
al.} (solid histogram).}
\label{f:kme:dlnu_w}
}
The results of the fit are given in table~\ref{t:kme:dlnu_ff}.  We
first parameterize the form factor as a Taylor expansion about $w=1$:
\begin{equation}
{\cal F}_D(w)={\cal F}_D(1) (1-\rho_D^2(w-1) +c_D(w-1)^2).  
\end{equation}
We first fit using only a linear term ($c_D=0$) and
then include the curvature term.
We also parameterize the form factor using the result of Boyd {\em et
al.} \cite{kme:boyd}, which uses dispersion relation to constrain
terms of higher order in $(w-1)$.  Similar results are obtained using the
parameterization of Caprini {\em et al.} \cite{kme:caprini96,kme:caprini98}.

We obtain the total decay rate for $B\to D\ell\nu$ by integrating
$d\Gamma/dw$ over $w$ using best fit values to Boyd {\em et al.}'s
parameterization of the the form factor.  We find
$\Gamma(B\to D\ell\nu)= (14.1 \pm 1.0 \pm 1.2)$ ns$^{-1}$,
where we have combined $B^-\to D^0\ell^-\bar\nu$ and ${\bar
B^0} \to D^+\ell-\bar\nu$ samples by assuming that $B^0\bar B^0$ and
$B^+B^-$ saturate the decays of the $\Upsilon(4S)$.  Using measured $B$
lifetimes this implies the branching fractions, 
${\cal B}(B^- \to D^0\ell^-{\bar\nu}) =  (2.32 \pm 0.17 \pm 0.20)$\%
and 
${\cal B}({\bar B^0} \to D^+\ell^-{\bar\nu}) = (2.20 \pm 0.16 \pm 0.19)$\%,
where the errors are statistical and systematic respectively.  Since
we derive the branching fractions from the decay width, the errors are
completely correlated.  We combine these results with the
previous CLEO measurement \cite{kme:Dlnu97} taking into account all
correlations:
\begin{eqnarray}
\Gamma(B\to D\ell\bar\nu) & = & (13.4 \pm 0.8 \pm 1.2 )\ {\rm ns}^{-1}
\nonumber \\
{\cal B}(B^- \to D^0\ell^-{\bar\nu}) & = & (2.21 \pm 0.13 \pm 0.19)\% 
\nonumber \\
{\cal B}({\bar B^0} \to D^+\ell^-{\bar\nu}) & = & (2.09 \pm 0.13 \pm 0.18)\% 
\nonumber
\end{eqnarray}
\begin{eqnarray}
{\cal F}_D(1) |V_{cb}| = (4.16 \pm 0.47 \pm 0.37) \times 10^{-2} \nonumber
\end{eqnarray}
Theoretical expectations for ${\cal F}_D(1)$ range from $0.98\pm0.07$
\cite{kme:caprini96} to $1.04$ \cite{kme:isgw1,kme:isgw2}.
A recent lattice calculation finds
$1.058^{+0.020}_{-0.017}$ \cite{kme:FDlatt2}.
Using ${\cal F}_D(1)=1.05\pm0.03$ we find 
\begin{eqnarray}
|V_{cb}|=0.040 \pm 0.004 \pm 0.004 \pm 0.001, \nonumber
\end{eqnarray}
where the errors are statistical, systematic and due to
theoretical uncertainty in ${\cal F}_D(1)$.  This value of $|V_{cb}|$
is consistent with that obtained in studies of the decay $B\to
D^*\ell\nu$.  If we use, instead, the best fit parameters to a linear
form factor, the value of $|V_{cb}|$ decreases by 10\%. This shows the
importance of the higher order terms in the form factor when
extrapolating to $w=1$.

\FIGURE{ \epsfig{file=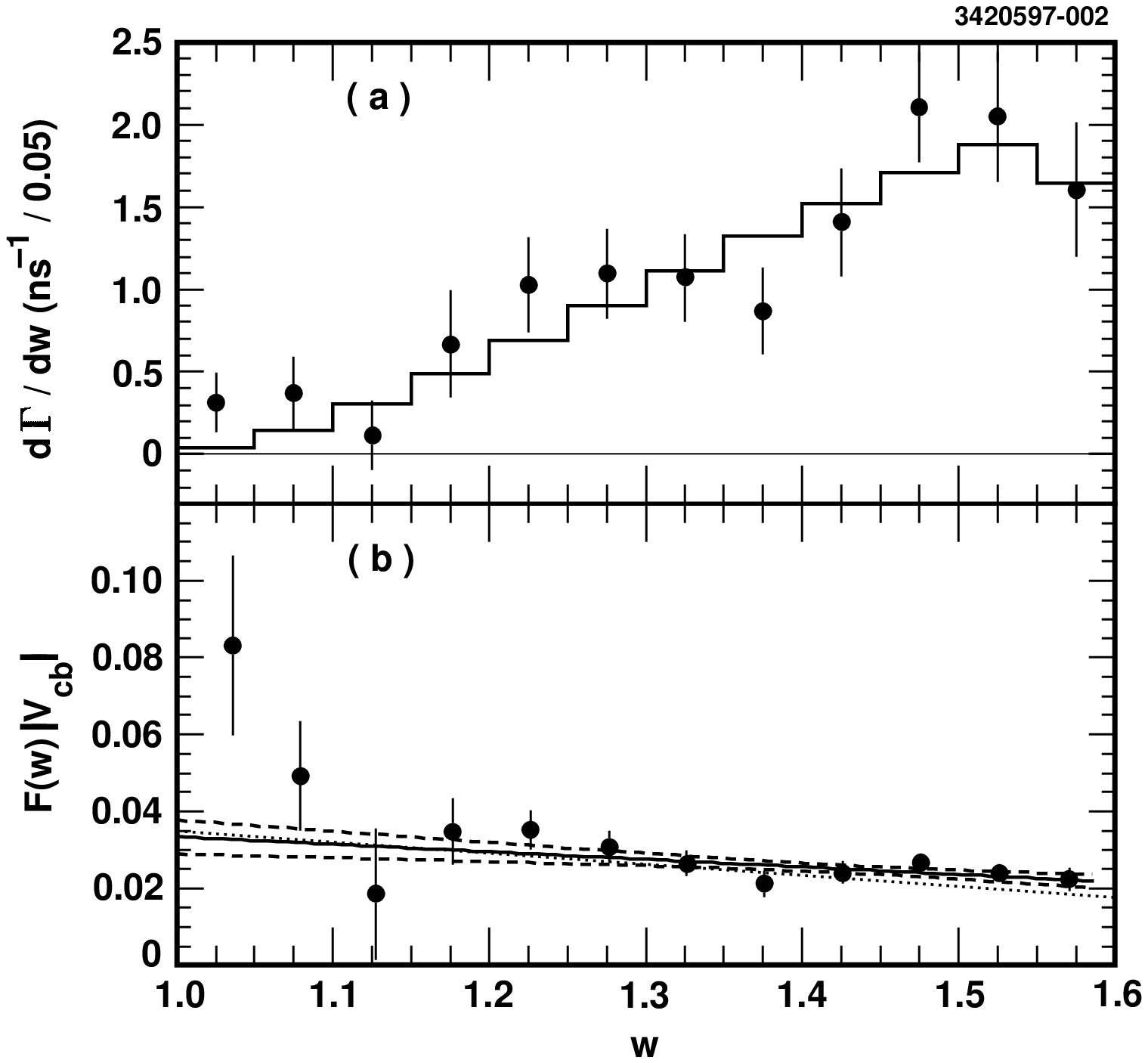,width=6.8cm}
\caption{Overlay of ${\cal F}_{D^{(*)}}(w)|V_{cb}|$ where the points
are $B\to D\ell\nu$ data~\protect\cite{kme:Dlnu97}, the solid line is the best
fit with statistical errors shown by the dashed lines, and the dotted
line shows the best fit from $B\to D^*\ell\nu$~\protect\cite{kme:Dslnu95}.}
\label{f:kme:dlnu_bloom}
}
The form factor for $D\ell\nu$ may also be compared with $D^*\ell\nu$.
In HQET at lowest order there is a common form factor.  Corrections at
higher order have also been calculated.  Figure~\ref{f:kme:dlnu_bloom}
plots ${\cal F}(w)|V_{cb}|$ vs $w$ for both $D\ell\nu$ and
$D^*\ell\nu$.  With the statistics available the agreement is
excellent as predicted by HQET.

\section{$B\to \pi\ell\nu$ and $B \to \rho\ell\nu$}
CLEO has also measured $b\to u\ell\nu$ decays which are sensitive to
$|V_{ub}|$.  Experimentally such measurements are difficult due to
large backgrounds from the Cabibbo favored $b\to c\ell\nu$ decays.

\subsection{1996 $B\to \pi\ell\nu$ and $B\to \rho\ell\nu$ Analysis}
\FIGURE{
\epsfig{file=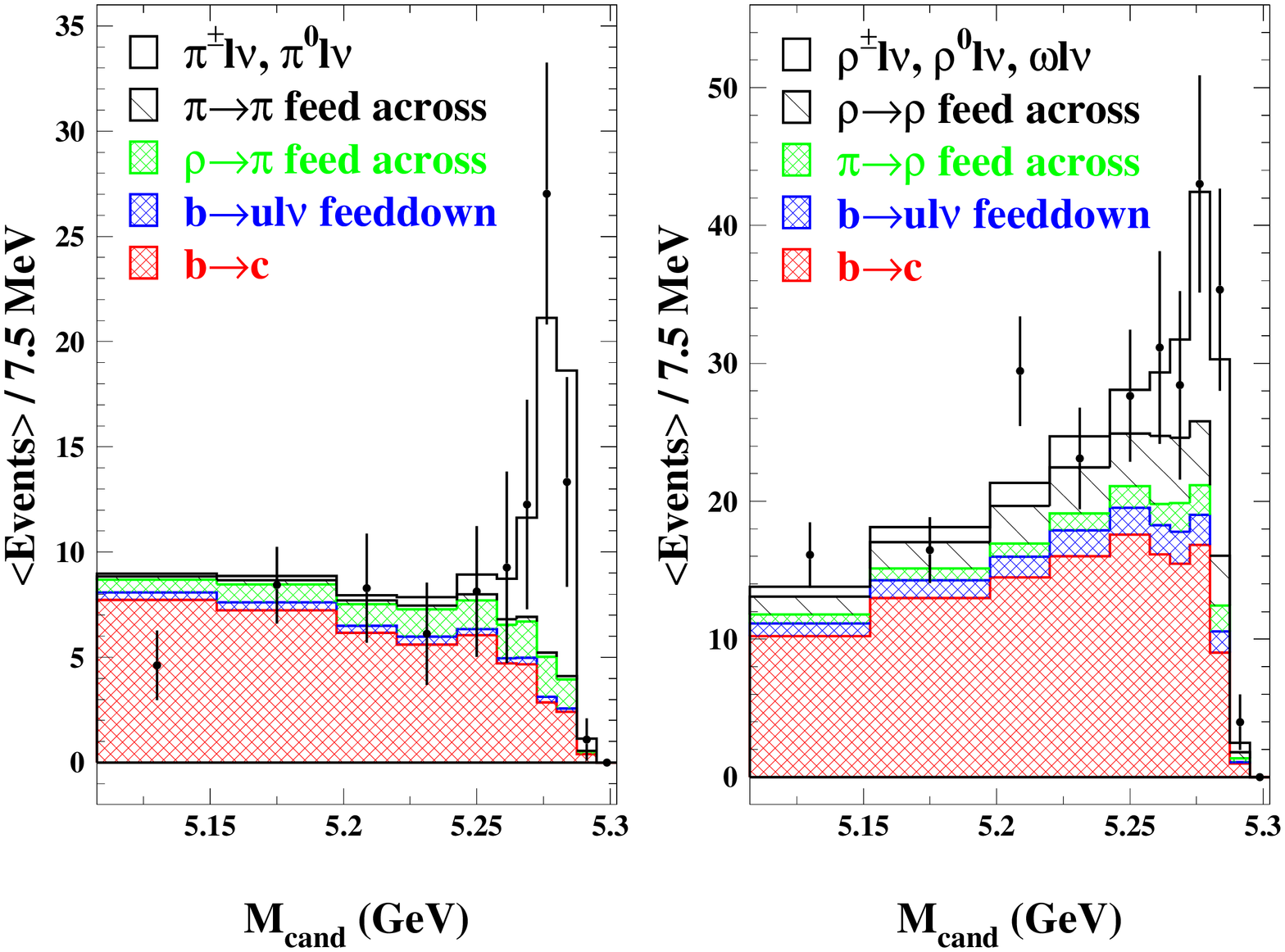,angle=-90,width=10cm}
\caption{$M_B$ distributions from 1996 analysis showing $\pi\ell\nu$
(left) and $\rho\ell\nu$ (right).}
\label{f:kme:pilnu96}
}
In 1996 CLEO observed the exclusive decay modes $B\to \pi \ell \nu$
and $B \to \rho (\omega) \ell \nu$ using the method of neutrino
reconstruction~\cite{kme:rholnu96}. Using the hermeticity of the CLEO
detector, the neutrino momentum and energy are estimated to be the
missing momentum and energy in the event.
\begin{eqnarray}
{\bf p}_\nu & = & -\sum_i {\bf p}_i \\
E_\nu & = & 2 E_{\rm beam} - \sum_i E_i
\end{eqnarray}
To ensure a good measurement of the missing energy and momentum, we
require the net charge of the event to vanish and carefully identify
showers in the crystal calorimeter to avoid double counting of
interacting charged particles.  To remove events with more than one
neutrino, we require that there be only one charged lepton identified
in the event.  The neutrino mass inferred from the missing energy and
momentum must be consistent with zero.  The resolution on the missing
energy (momentum) is 260 MeV (110 MeV/$c$).

Electrons are identified using the CSI calorimeter, and muons are
identified by hits in the muon counters at depths of 5 and 7
interaction lengths.  Backgrounds from $b\to c\ell\nu$ and $b\to c \to
s\ell\nu$ are significantly reduced by requiring $p_{\ell} > 1.5 (2.0)
$ GeV/$c$ for the $\pi$ ($\rho$) mode.  The $\rho^\pm$ and $\omega$
candidates are identified in the $\pi\pi$ and $\pi^+\pi^-\pi^0$ decay
modes respectively. 

The reconstructed neutrino 4-vector ($|{\bf p}_\nu|,{\bf p}_\nu$) is
combined with a lepton and $\pi$ or $\rho$ candidate to form $B$
candidates, which will have the $B$ meson mass and beam energy for
signal events.  The signal is extracted by fitting the variables $M_B$
and $\Delta E$.
\begin{eqnarray}
\Delta E & = & E_{\pi(\rho)} + E_\ell + |{\bf p}_\nu| - E_{\rm beam} \\
M_B &=& \sqrt{E^2_{\rm beam}-|{\bf p}_{\pi(\rho)} + {\bf p}_\ell + {\bf
p}_\nu |^2} \nonumber
\end{eqnarray}
For $\rho\ell\nu$ we fit $M(\pi\pi)$ as well.  The fit determines
components for signal and background from $b\to c$.  Isospin and quark
model relations are used to fix the relative rate of $B^0$ and $B^+$
decays.  We take $\Gamma(B^0 \to \pi^-\ell^+\nu) = 2\Gamma(B^+\to
\pi^0\ell^+\nu)$ and $\Gamma(B^0 \to \rho^-\ell^+\nu) = 2\Gamma(B^+\to
\rho^0\ell^+\nu) \approx 2\Gamma(B^+\to\omega\ell^+\nu)$, leaving two
independent yields $N_\pi$ and $N_\rho$.  Cross-feed among the different
modes and from $B\to X_u\ell\nu$ (higher resonances) is included in
the fit.  The fit result is shown in figure~\ref{f:kme:pilnu96}.

We see clear signals in $B\to\pi\ell\nu$ and $B\to\rho\ell\nu$.  The
yield in $B\to\omega\ell\nu$ is consistent both with the expectations
from $\rho\ell\nu$ and with background.
The branching fractions for the observed signals are determined using
efficiencies from a full detector MC of the signals for five form
factor models \cite{kme:isgw2,kme:wsb,kme:ks,kme:melikhov}.
Each model independently predicts the $\pi/\rho$ ratio.  We compare
this to the data, and for four of the five models, the ratio is
consistent.  The remaining model is excluded, and we average the
results from others.
\begin{eqnarray}
{\cal B}(B^0 \to \rho^- \ell^+ \nu) & = & 
(2.5 \pm 0.4 ^{+0.5}_{-0.7} \pm 0.5) \times 10^{-4} \nonumber \\
{\cal B}(B^0 \to \pi^- \ell^+ \nu) & = & 
(1.8 \pm 0.4 \pm 0.3 \pm 0.2) \times 10^{-4} \nonumber \\
|V_{ub}| &=& (3.3 \pm 0.2 ^{+0.3}_{-0.4}\pm 0.7) \times 10^{-3} \nonumber
\end{eqnarray}
The uncertainties on the measurements are statistical, systematic and
due to model dependence.  The model dependence is estimated from the
spread of the results using different models.  This uncertainty
($\sim 20$ \%) currently limits the accuracy of $|V_{ub}|$ measurements
using exclusive $b\to u\ell\nu$ decays.  However, with more $B$ decays,
one can begin to differentiate between the different models.  The aim
of a second CLEO analysis is to increase the efficiency and begin to
do just that by measuring the $q^2$ dependence.

\subsection{1999 $B\to \rho\ell\nu$ Analysis}
In this second analysis \cite{kme:rholnu}, we analyze the decay $B\to
\rho\ell\nu$ using high momentum leptons paired with $\pi,\rho$ and
$\omega$ candidates.  In the high momentum region we are able to
measure the $q^2$ distribution of $B\to\rho\ell\nu$ events.

We select events with leptons of energy  $E_\ell>1.7$ GeV/$c$
accompanied by a hadronic system consistent with a $\rho$
($\pi^+\pi^-$ or $\pi^\pm\pi^0$), $\omega$ ($\pi^+\pi^-\pi^0$) or
$\pi$ ($\pi^\pm$ or $\pi^0$). To reduce background from $b\to c\ell\nu$
decays we divide the sample into three lepton energy bins:
HILEP   (2.3--2.7 GeV/$c$),
LOLEP   (2.0--2.3 GeV/$c$) and
LOLOLEP (1.7--2.0 GeV/$c$).
Leptons in the HILEP bin have energy above the kinematic endpoint for
$b\to c\ell\nu$ decays.  The LOLEP bin contains mostly $b\to c
\ell\nu$ events but still has some sensitivity to $b\to u \ell\nu$.
The lowest energy bin provides a normalization of the $b\to c
\ell\nu$ background.

The dominant source of background in the highest energy bin comes
from continuum production of hadrons: $e^+e^- \to q \bar q$,
$q=u,d,s,c$.  Since the decays of $B$ mesons at rest are more
spherical than jet-like $q\bar q$ events, we suppress this background
using event shape variables.  We obtain additional suppression by
requiring $\cos\theta_{B-\rho\ell}$ to be physical. (See
equation~\ref{e:kme:cby}.)

For each $\rho\ell\nu$ candidate, we compute 
$\Delta E = E_\rho + E_\ell + |{\bf p}_{\rm miss}| - E_{\rm beam}$,
where ${{\bf p}_{\rm miss}}$ is the net missing momentum in the event.
For signal events, $\Delta E$ should peak near zero since ${\bf p}_{\rm
miss}$ gives a measure of the neutrino energy and momentum.  Because
we rely on the hermeticity of the detector for this measurement, we
require the missing momentum not to point down the beam pipe.  We also
require ${\bf p}_{\rm miss}$ to be within 35$^\circ$ of the neutrino
direction inferred from the $\rho + \ell$ candidate; the later is
known up to an azimuthal ambiguity about the $B$ momentum direction.

\FIGURE{
\epsfig{file=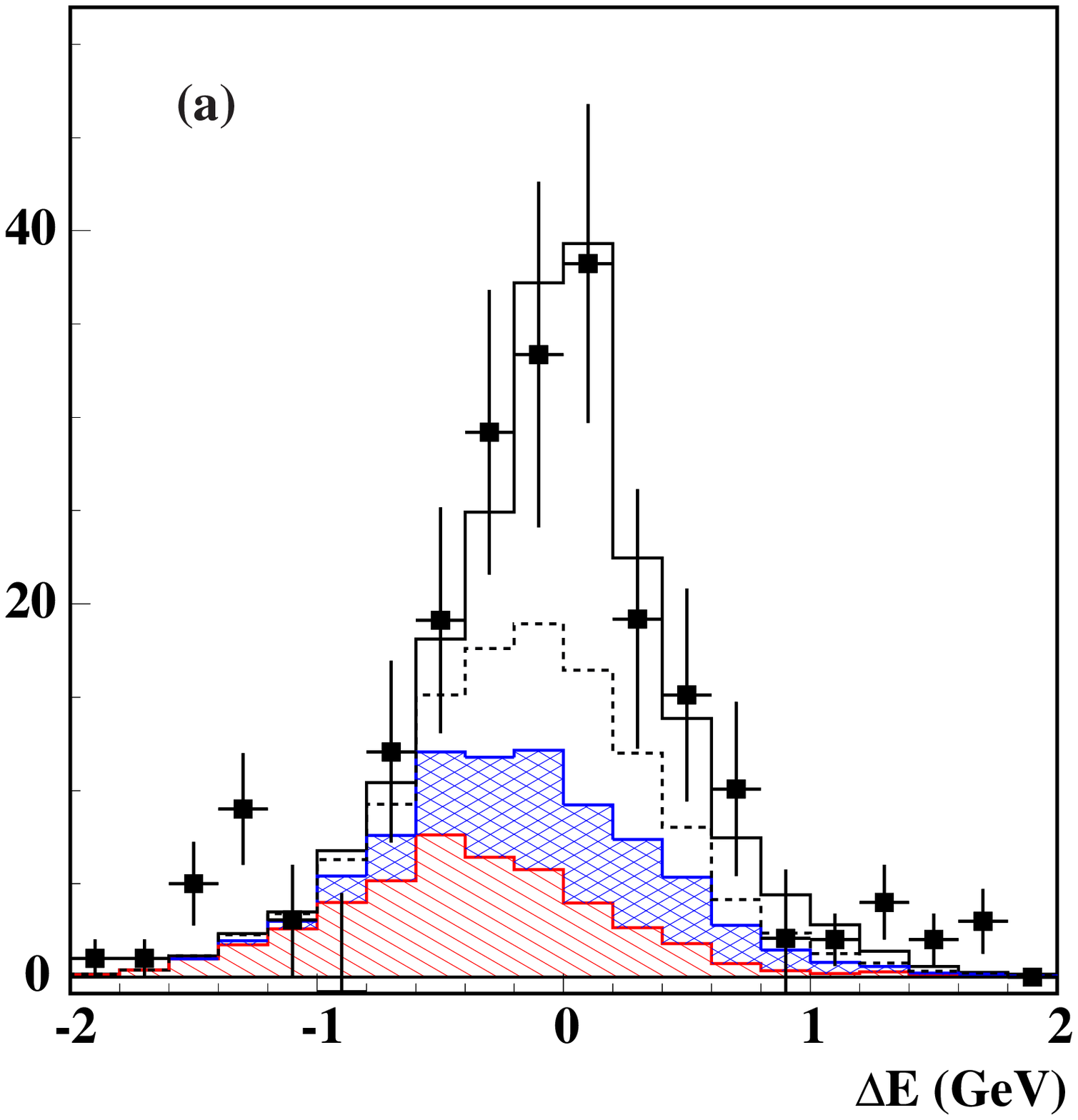,width=2.5in}
\epsfig{file=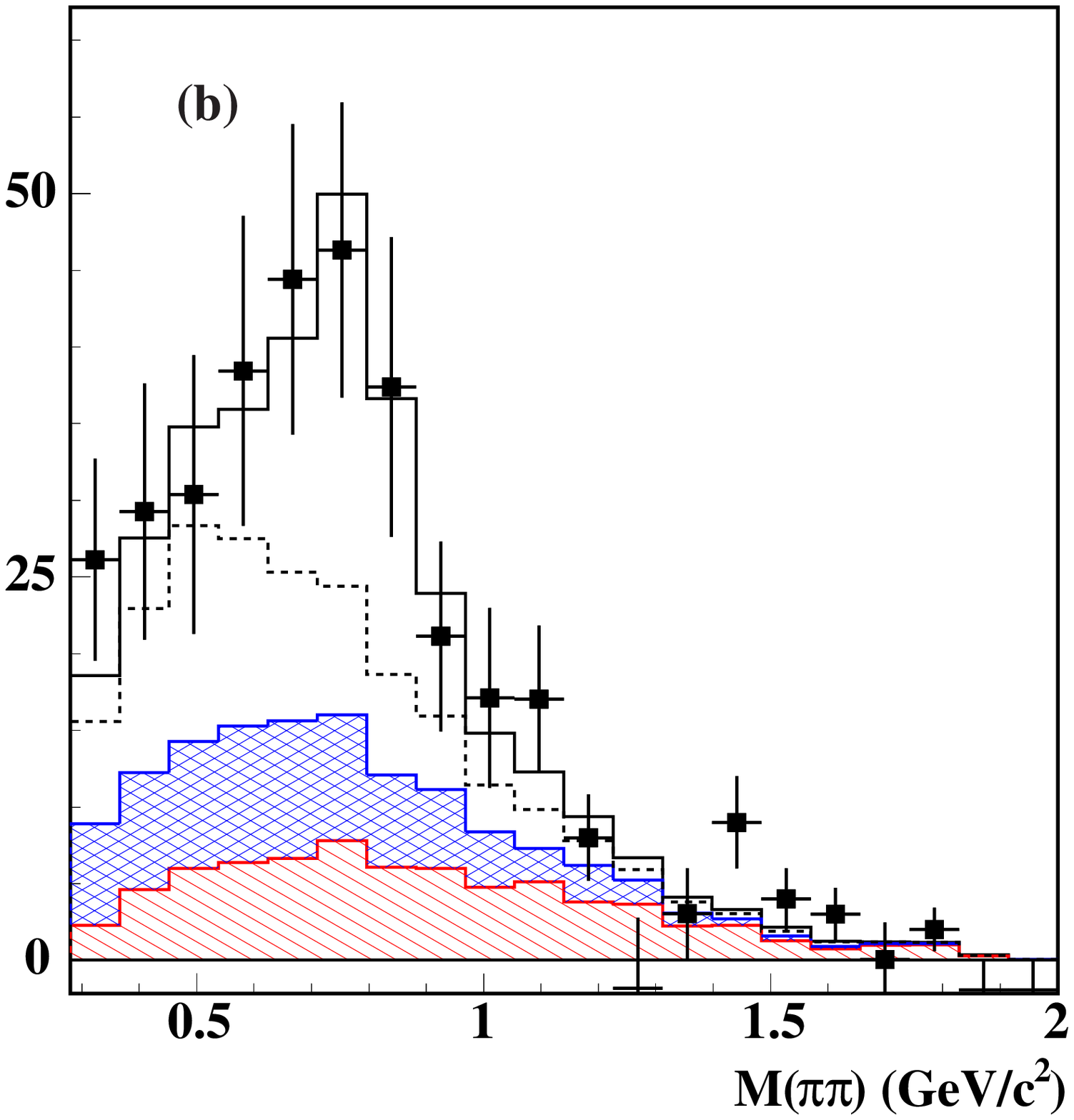,width=2.5in}
\caption{Projections of maximum likelihood fit for HILEP energy bin:
(a) $\Delta E$ distribution with the cut
$|M(\pi\pi)-M_\rho|<0.15$ GeV/$c^2$ and 
(b) $M(\pi\pi)$ distribution with the cut $\Delta E<500$ MeV.  The
points are the continuum subtracted data.  The solid histogram is the
fit, represented as the sum of three components: signal and cross-feed
(open regions), background from non-signal $b\to u\ell\nu$
(double-hatched region) and  background from $b\to c\ell\nu$
(single-hatched region).}
\label{f:kme:rholnu_proj}
}

To measure the $\rho\ell\nu$ branching fraction, we perform a
simultaneous maximum likelihood fit for all five modes in all three
lepton energy bins.  We fit in two variables, $\Delta E$ and
$m(\pi\pi(\pi))$, for the $\rho$ and $\omega$ modes; for the $\pi$
modes, we fit only to $\Delta E$.  The fit contains contributions from
the physics processes $B\to\rho(\omega)\ell\nu$, $B\to\pi\ell\nu$,
$b\to u\ell\nu$ (modes other than $\rho,\omega$ and $\pi$) and $b\to
c\ell\nu$. The fit also contains background contributions from
continuum and fake leptons; we measure these contributions using
off-resonance data and known fake rates.  The signal shapes for the
fit are taken from Monte Carlo simulation using various form factor
models for $B\to\rho\ell\nu$ and $B\to\pi\ell\nu$
\cite{kme:isgw2,kme:lcqcd,kme:beyer,kme:ukqcd,kme:e791}, the ISGW2
\cite{kme:isgw2} model for $b\to u\ell\nu$ and a combination of ISGW2
and CLEO form factor results \cite{kme:Dslnu95} for $b\to c\ell\nu$.
Isospin and quark model relations are used to constrain the relative
normalizations of the three vector modes ($B^0\to \rho^-\ell^+\nu$,
$B^+\to \rho^0\ell^+\nu$ and $B^+\to\omega\ell^+\nu$) and, separately,
the normalizations of the pseudoscalar modes ($B^0\to \pi^-\ell^+\nu$
and $B^+\to\pi^0\ell^+\nu$).  Our fit also accounts for the large
cross-feed between the various signal modes.

We find statistically significant yields for $B\to\rho\ell\nu$.
Figure~\ref{f:kme:rholnu_proj} shows projections of the maximum
likelihood fit for $\pi^+\pi^-$ and $\pi^\pm\pi^0$ modes in the high
energy bin onto the variables $\Delta E$ and $M(\pi\pi)$ overlayed
with the data.
We average over the various form factor models for $\rho\ell\nu$ and
$\pi\ell\nu$, finding
\begin{eqnarray}
{\cal B}(B\to \rho\ell\nu) &=& 
(2.69 \pm 0.41^ {+0.35}_{-0.40} \pm 0.50) \times
10^{-4} \nonumber \\
|V_{ub}|&=&
(3.23 \pm 0.24^{+0.23}_{-0.26} \pm 0.58)\times 10^{-3}, \nonumber
\end{eqnarray}
where the errors are statistical, systematic and due to model
dependence.  These results for ${\cal B}(B\to\rho\ell\nu)$ and $|V_{ub}|$
are consistent with the neutrino reconstruction analysis
\cite{kme:rholnu96}.  The two results are statistically independent,
but the systematic and model dependence uncertainties are largely
correlated.  Taking into account the correlations, the combined results
are
\begin{eqnarray}
{\cal B}(B^0\to\rho^-\ell^+\nu) & = & 
(2.57 \pm 0.29 ^{+0.33}_{-0.46}\pm 0.41) \times 10^{-4} \nonumber \\
|V_{ub}| & = & 
(3.25 \pm 0.14 ^{+0.21}_{-0.29} \pm 0.55) \times 10^{-3} \nonumber
\end{eqnarray}
The $\pi\ell\nu$ mode is dominated by cross-feed from
$\rho\ell\nu$, but the branching fraction ${\cal
B}(B\to\pi\ell\nu)=(1.3\pm0.4)\times 10^{-4}$ (statistical error only)
is consistent with the neutrino reconstruction analysis.  In
$\omega\ell\nu$, the fit describes the data well but we do not observe
a significant signal.

\FIGURE{
\epsfig{file=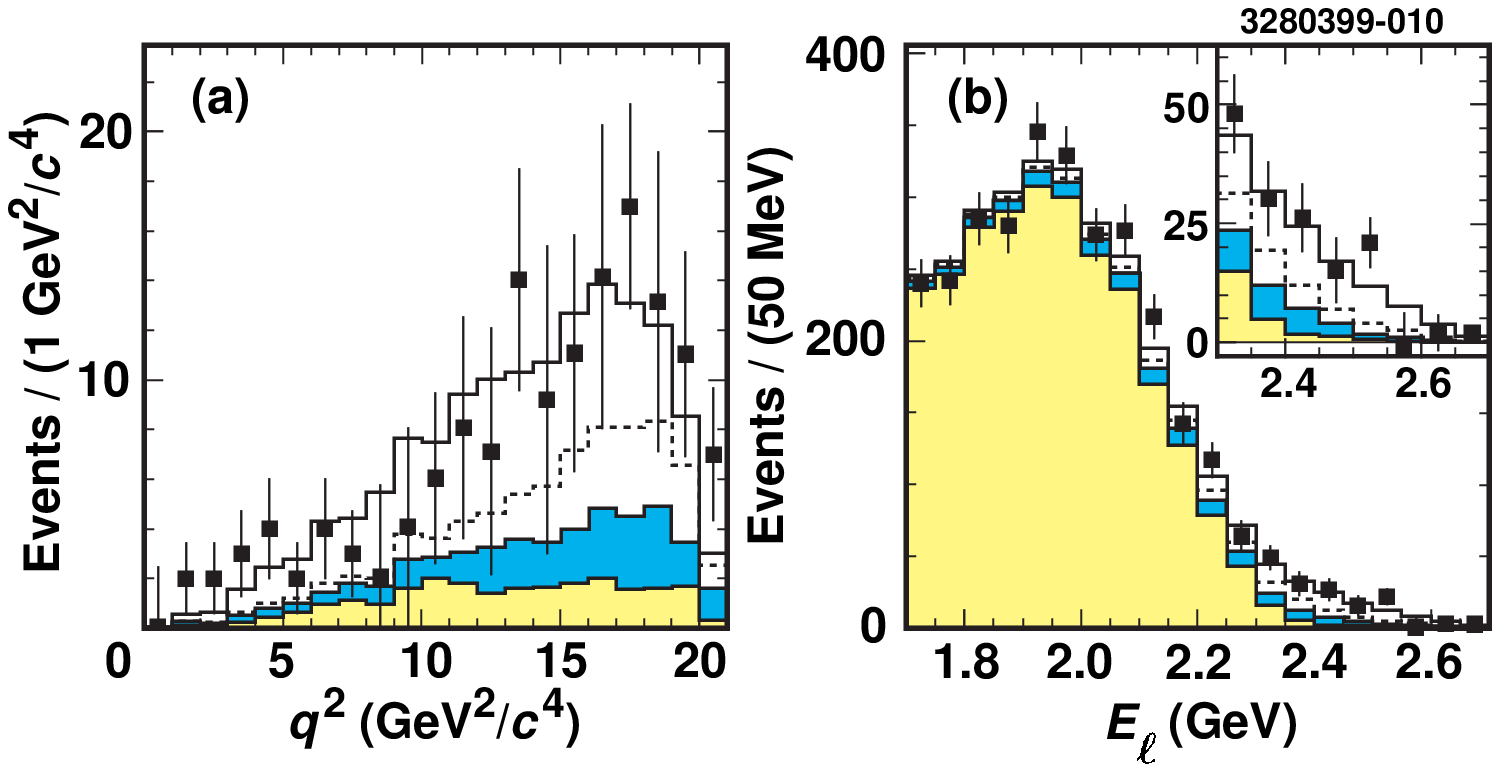,width=\textwidth}
\caption{The projections of the fit onto $q^2$ (a) and $E_\ell$ (b)
for HILEP after the cuts $\Delta E<500$ MeV and
$|M(\pi\pi)-M_\rho|<0.15$ GeV/$c^2$.  The points show the on-resonance
data after continuum subtraction, while the histogram shows the
projection of the fit.  The contributions to the fit are direct and
cross-feed components of the signal (unshaded regions above and below
the dashed line), the background from $b\to u\ell\nu$ non-signal modes
(darkly shaded region) and background from $b\to c\ell\nu$ (lightly
shaded region).
}
\label{f:kme:rholnu_q2}
}
We are also able to measure the $q^2$ distribution for
$B\to\rho\ell\nu$ events with $E_\ell>2.3$ GeV.
Figure~\ref{f:kme:rholnu_q2} shows the data distribution of $q^2$
after requiring $|M(\pi\pi)-M_\rho|<0.15$ GeV/$c^2$ and $\Delta E< 500$
MeV.  We quote partial widths for three $q^2$ bins in
table~\ref{t:kme:DeltaG_rholnu}.
\TABLE{
\caption{The partial width $\Delta\Gamma$ for $B\to\rho\ell\nu$ in
bins of $q^2$. }
\label{t:kme:DeltaG_rholnu}
\begin{tabular}{rcl}
$\Delta\Gamma(0 < q^2 <7$ GeV$^2/c^4)$ & = &
  $(7.6 \pm 3.0^{+0.9}_{-1.2} \pm 3.0) \times 10^{-2} \ {\rm ns}^{-1}$ \\
$\Delta\Gamma(7 < q^2 < 14$ GeV$^2/c^4)$ & = &
  $(4.8 \pm 2.9^{+0.7}_{-0.8} \pm 0.7) \times 10^{-2} \ {\rm ns}^{-1}$ \\
$\Delta\Gamma(14 < q^2 < 21$ GeV$^2/c^4)$ & = &
  $(7.1 \pm 2.1^{+0.9}_{-1.1} \pm 0.6) \times 10^{-2} \ {\rm ns}^{-1}$ 
\end{tabular}
}
The measurements are subject to statistical, systematic and model
dependence uncertainties.  The model dependence uncertainty comes
primarily from the extrapolation to all lepton energies.  In the
highest $q^2$ bin, the model dependence is small because the
experimentally accessible lepton energies ($E_\ell>2.3$ GeV) cover
fractionally more of the allowed phase space.  We compare the measured
differential decay rate to expectations from different form factor
models in figure~\ref{f:kme:rholnu_dg}.
\FIGURE{
\epsfig{file=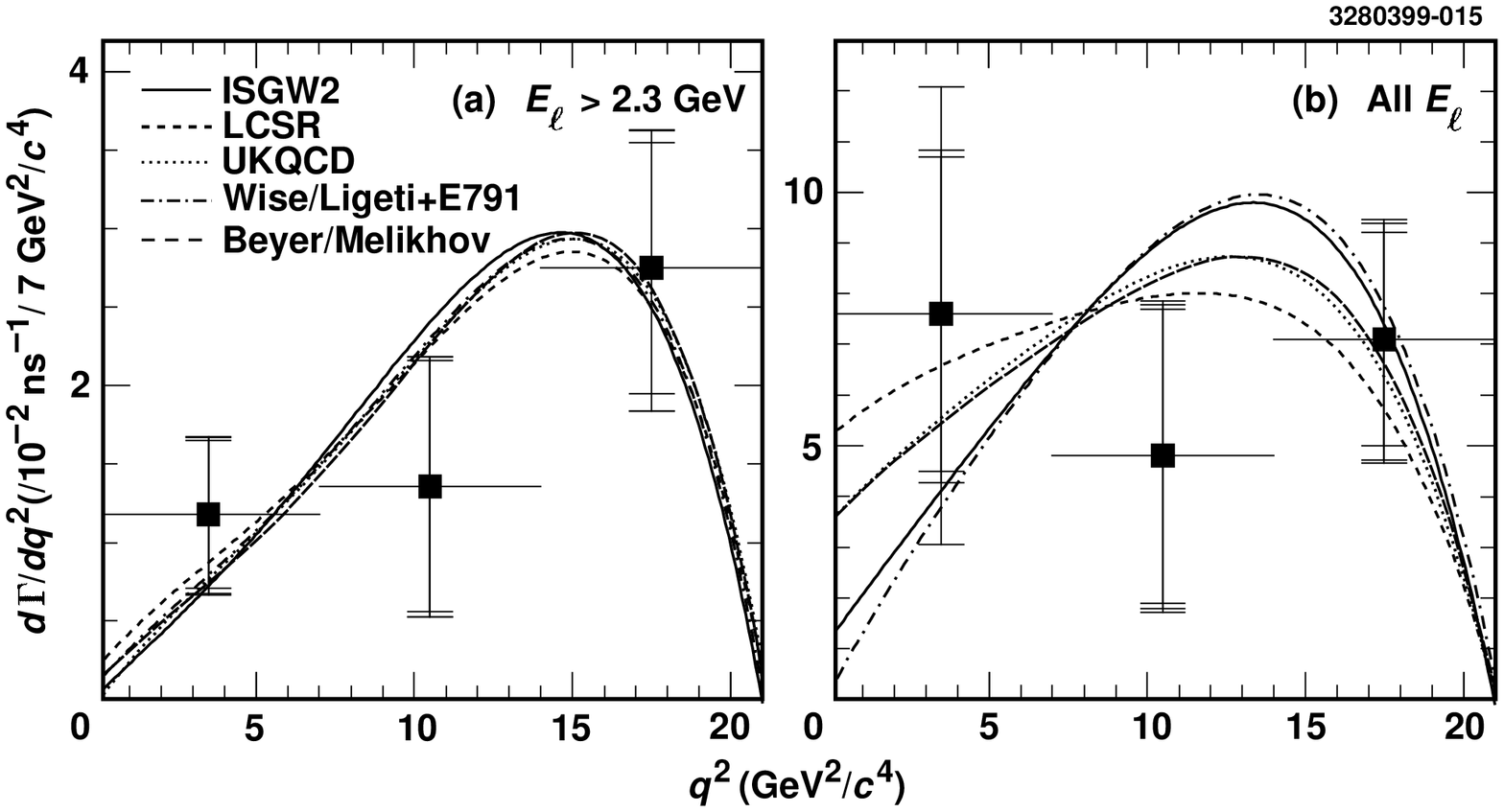,width=6.8cm,angle=-90}
\caption{Comparison of measured differential rate $d\Gamma/dq^2$ in
three bins and expectations from five form factor models.  (a) shows
the results for $E_\ell>2.3$ GeV/$c$, while (b) extrapolates the data
to the entire lepton energy range.}
\label{f:kme:rholnu_dg}
}
At high lepton energy, the form factor models agree well on the shape of
the form factor, which is dominated by one of three relevant form
factor terms ($A_1(q^2)$), and in fact the model dependence might be
quite small.
In a future analysis, one might choose to measure $|V_{ub}|$ using the
decay rate at large $q^2$.  At the same time, the good agreement
removes the possibility of differentiating between models
at this time.  Measurement of the partial rate for $E_\ell < 2.0$ GeV
would help improve the form factor models, and thus improve
measurement of $|V_{ub}|$.  It is also possible lattice QCD
calculations can provide more precise information about the form factor
in an experimentally accessible region of $q^2$ and $E_\ell$.

\section{Analysis of Inclusive $B\to X_c\ell\nu$}
Inclusive measurements of $b\to c\ell\nu$ also give information on
$|V_{cb}|$.  CLEO has two preliminary results based on inclusive
techniques for measuring semileptonic decays
\cite{kme:moments,kme:ser}.  Both rely on Heavy Quark Effective Theory
(HQET) and the operator product expansion (OPE) to interpret the
results.  Within this framework an inclusive measurement summed over
many final states is readily interpreted from quark level
calculations.  

The rate for inclusive $B\to X_c \ell\nu$ is proportional to
$|V_{cb}|^2$.  In the OPE and HQET to order $1/M_b^2$ the rate may be
written \cite{kme:fls,kme:fl,kme:gremm}
\begin{eqnarray}
\Gamma_{sl}&=&{G_F^2 |V_{cb}|^2 M_B^5 \over 192 \pi^3} 
    0.369 \bigg[1-1.54 {\alpha_s \over \pi} \\
&& -1.65 {\bar \Lambda \over M_B} (1-0.87 {\alpha_s \over \pi}) 
   -0.95 {\bar \Lambda^2 \over M_B^2} \nonumber \\
&& -3.18 {\lambda_1 \over M_B^2} 
   +0.02 {\lambda_2 \over M_B^2} \bigg]
\nonumber 
\end{eqnarray}
The parameters $\lambda_1$ and $\lambda_2$ are matrix elements of the
HQET expansion, which have the following intuitive interpretations:
$\lambda_1$ is proportional to the kinetic energy of the $b$-quark in
the $B$ meson and $\lambda_2$ is the energy of the hyperfine
interaction of the $b$-quark spin and the light degrees of freedom in
the meson.  $\bar\Lambda$ is introduced to relate the $b$-quark and
$B$ meson masses, representing the energy of the light degrees of
freedom.

From the $B$-$B^*$ mass difference, $\lambda_2$ is determined to be
0.12 GeV$^2$.  $\bar\Lambda$ and $\lambda_1$ are more difficult to
determine, but if they can be measured, one can measure $|V_{cb}|$
given $\Gamma_{sl}$.  For example, from 
${\cal B}(B\to Xe\nu)=(10.49\pm 0.17\pm 0.43)$\% \cite{kme:cleolep}
and the average $B$ lifetime $\tau_B=1.61 \pm 0.02$ ps one finds
$\Gamma_{sl}=65.0 \pm 3.0$ ns$^{-1}$.
At present our knowledge of $\lambda_1$ and $\bar\Lambda$
limits the precision we can achieve on $|V_{cb}|$ from inclusive
semileptonic $B$ decays.  The aim of the new inclusive analyses is to
determine $\lambda_1$ and $\bar \Lambda$ from experiment and thereby
decrease the theoretical uncertainty which comes when extracting
$|V_{cb}|$ from $\Gamma_{sl}$.  Each analysis alone provides two
constraints, allowing a measurement of $\bar\Lambda$ and $\lambda_1$.
Combining the two analyses over-constrains the theory parameters thus
allowing a test of the theoretical framework and experimental
understanding of $b$-quark decays.

\subsection{Hadronic Mass Moments}
\FIGURE{
\epsfig{file=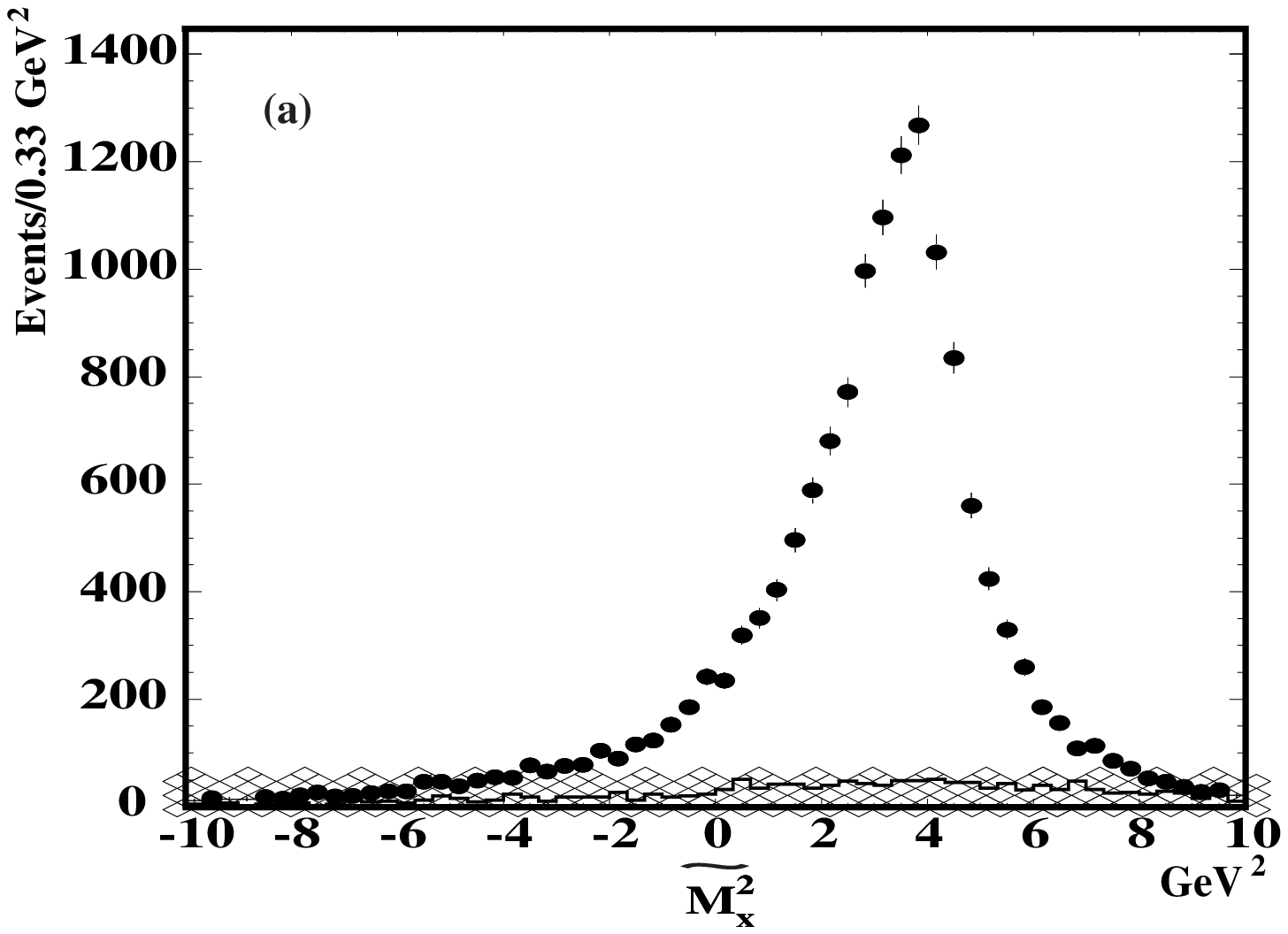,width=2.8in}
\epsfig{file=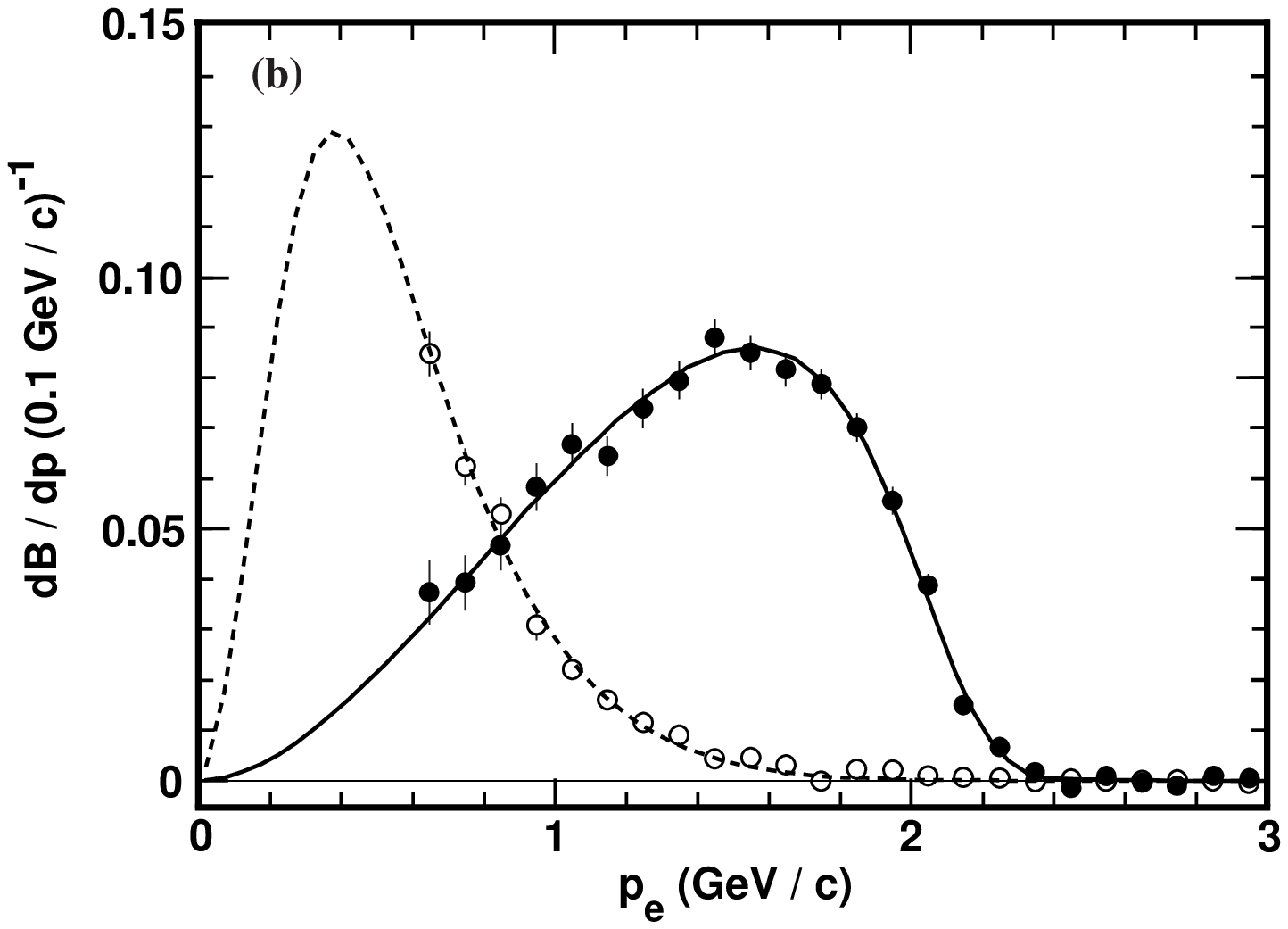,width=2.8in}
\caption{
(a) Measured $\tilde{M}^2_{X_c}$ distributions for
on-resonance data (points) and scaled off-resonance data (hatched
histogram).
(b) Electron momentum spectrum from $B\to X e\nu$ (solid circles)
and $b\to c \to Y e\nu$ (open circles).  The curves show the best fit
to the ISGW model with 23\% $B\to D^{**}\ell\nu$.
}
\label{f:kme:hadmom}
\label{f:kme:lepspec}
}

For decays $B\to X_c\ell\nu$, the first method measures the first and
second hadronic mass moments.  Falk {\em et al.}~\cite{kme:fls} give
an expansion for the moments of the squared hadronic invariant mass
($M^2_{X_c}$) distribution in the variables $1/M_B$ and $\alpha_s$.
The moments have been calculated integrated over all lepton energies
\cite{kme:fls,kme:gremm} and subject to a necessary experimental cut on the
lepton energy \cite{kme:fl}.  Equations~\ref{e:kme:hadmom} and
\ref{e:kme:hadmom2} give the expansions for the first and
second moments to order $1/M_B^2$, for $E_\ell>1.5$ GeV.
\begin{eqnarray}
\langle M_{X_c}^2 &-& \bar M_D^2 \rangle =
 M_B^2 \ \bigg[  0.0272 \frac{\alpha_s}{\pi}  \label{e:kme:hadmom}\\
&      + & 0.207  \frac{\bar \Lambda}{M_B}(1+0.43 \frac{\alpha_s}{\pi})
 \nonumber  \\
&      + & 0.193  \frac{\bar \Lambda^2}{M_B^2}  
         + 1.38   \frac{\lambda_1}{M_B^2} 
         + 0.203  \frac{\lambda_2}{M_B^2}\bigg] \nonumber
\end{eqnarray}
\begin{eqnarray}
\langle (M_X^2 &-&\bar M_D^2)^2 \rangle = M_B^4\bigg[ 0.00148
  \frac{\alpha_s}{\pi}  +  \label{e:kme:hadmom2}\\
& + & 0.038 \frac{\bar \Lambda}{M_B}\frac{\alpha_s}{\pi} 
  +   0.0535 \frac{\bar \Lambda^2}{M_B^2}
  -   0.12 \frac{\lambda_1}{M_B^2}\bigg] \nonumber
\end{eqnarray}
The moments are defined relative to the spin-averaged $D$ meson mass,
$\bar M_D=1.975$ GeV/$c^2$.
By measuring the first two moments and inverting the equations one may
determine or constrain the remaining HQET parameters $\lambda_1$ and
$\bar\Lambda$.

To measure the hadronic mass moments in semileptonic $B$ decays we
select events with one lepton of momentum $p_\ell > 1.5$ GeV/$c$.  We
``reconstruct'' the neutrino using the hermeticity of the detector,
imposing strict event quality cuts to ensure no particles are missed.
The net charge of the event must be zero, and the missing mass must be
consistent with a neutrino.  The mass recoiling against the lepton and
neutrino is:
\begin{eqnarray}
M_{X_c}^2 & = & M_B^2 + M_{\ell \nu}^2 - 2 E_B E_{\ell \nu} \\
&& + 2 \vert {\bf p}_B \vert \vert {\bf p}_{\ell \nu} \vert \cos
\theta_{\ell \nu-B}. \nonumber
\end{eqnarray}
Since the $B$ momentum is small but the direction is unknown, we 
approximate $M_{X_c}^2$ by dropping the last term.
\begin{equation}
\widetilde {M_{X_c}^2} = M_B^2 + M_{\ell \nu}^2 - 2 E_B E_{\ell \nu}
\end{equation}

The resulting distribution shown in figure~\ref{f:kme:hadmom}a has
contributions from $b\to c\ell\nu$ (96\%), $b\to c \to s\ell\nu$ (3\%)
and $b\to u\ell\nu$ (1\%).  We compute the moments after background
subtraction using MC shapes.  We further correct for a bias in the
reconstructed hadronic mass due to asymmetric resolution of the
neutrino reconstruction.  We find 
\begin{eqnarray}
\langle M_{X_c}^2 -\bar M_D^2 \rangle
& = & 0.286\pm0.023\pm0.080 \ {\rm GeV}^2 \nonumber \\
\langle (M_{X_c}^2 -\bar M_D^2)^2 \rangle 
& = & 0.911\pm0.066\pm0.309 \ {\rm GeV}^4. \nonumber
\end{eqnarray}

\subsection{Lepton Energy Moments}
The second method uses the inclusive electron spectrum from
$B$ decays measured by CLEO \cite{kme:cleolep}.  Theoretical
expressions for the moments of the lepton spectrum are given by
Voloshin \cite{kme:voloshin}.  As in the case of the
hadronic mass moments, these expressions may be inverted to place
constraints on $\lambda_1$ and $\bar\Lambda$.

The $B\to X e\nu$ electron spectrum measurement \cite{kme:cleolep}
shown in figure~\ref{f:kme:lepspec}b is an observed spectrum above 0.6
GeV.  In events with a high momentum lepton tag and an additional
electron, the primary electrons ($b\to c \ell^- X$) are separated from
secondary electrons from charm decays ($b\to c X; c \to \ell^+ Y$ )
using angular and charge correlations.  To measure the moments and
compare to theory, we must apply corrections to the observed primary
spectrum.  We extrapolate below 0.6 GeV and correct for detector
smearing (including bremsstrahlung) and motion of the $B$ in the lab
frame.  There are also electromagnetic radiative corrections which are
not included in the theoretical expressions for the moments.  After
all corrections we find the following preliminary results.
\begin{eqnarray}
\langle E_\ell \rangle &=& 1.36 \pm 0.01 \pm 0.02\ {\rm GeV}
\nonumber \\
\langle (E_\ell - \langle E_\ell \rangle )^2 \rangle &=& 0.19 \pm \nonumber
0.004 \pm 0.005\ {\rm GeV}^2
\end{eqnarray}

\FIGURE{
\epsfig{file=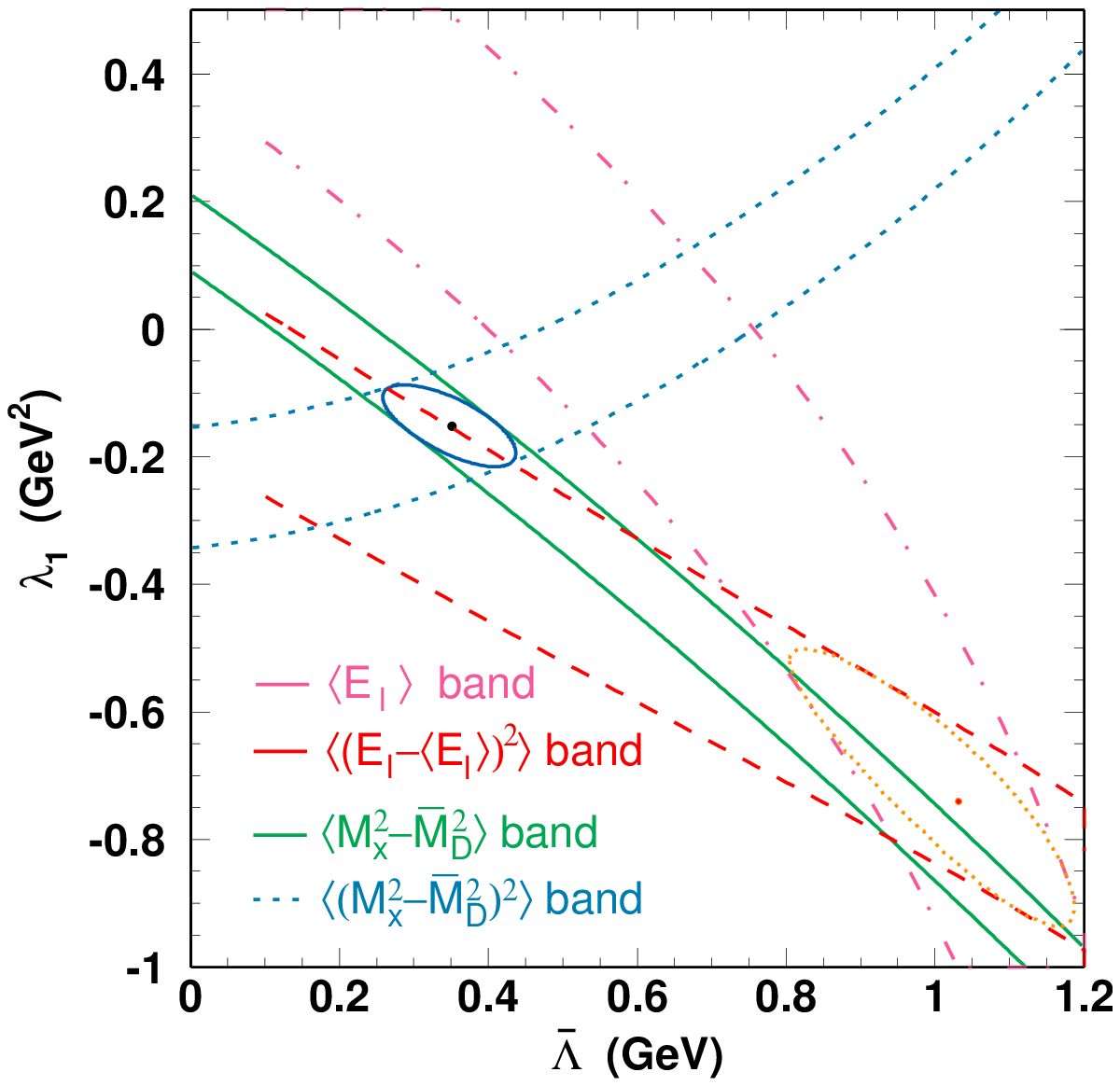,width=9cm}
\caption{Combined constraints on HQET parameters $\lambda_1$ and
$\bar\Lambda$ from hadronic recoil mass moments and lepton energy
spectrum moments.  
}
\label{f:kme:con}
}
\subsection{Interpretation of Moments}

Inverting the theoretical expression for the hadronic moments to 
${\cal O}(1/M_B^2)$ and solving for HQET parameters gives
the results
\begin{eqnarray}
\bar \Lambda &=& +0.33 \pm 0.02 \pm 0.08\ \ {\rm GeV}   \nonumber \\
   \lambda_1 &=& -0.13 \pm 0.01 \pm 0.06\ \ ({\rm GeV}/c)^2. \nonumber
\end{eqnarray}
Equivalently, each moment measurement provides a constraint in the
$\lambda_1$-$\bar\Lambda$ plane.  The allowed bands and overlap region
are shown in figure~\ref{f:kme:con}.
The 1 $\sigma$ allowed regions shown in the figure include statistical
and {\em both} experimental and theoretical systematic uncertainties.  We
use the ${\cal O}(1/M^3_B)$ expansion of the moments \cite{kme:fl} to
estimate the effect of higher order terms in the hadronic moments
calculation. Variations of the 1 $\sigma$ contours shown include this
theoretical systematic uncertainty.

The lepton moment measurements can also be converted to allowed bands
in the $\lambda_1$-$\bar\Lambda$ plane (figure~\ref{f:kme:con}).  In
these preliminary analyses, the agreement among the four allowed bands
is only at the 5--10\% confidence level.  Taken at face value, the
hadron moment measurement alone implies a 3\% measurement of
$|V_{cb}|$.  However, if one uses the central value from the lepton
moments instead, $|V_{cb}|$ shifts by $\sim 10$\%. Clearly before we
can feel comfortable with {\em precision} determinations of $|V_{cb}|$
or $|V_{ub}|$ from inclusive measurements, we must resolve the
discrepancy.

A few comments on the current discrepancy are in order.  First, Falk
{\em et al.} expect the second order hadronic mass moments to be more
sensitive to higher order corrections and therefore less reliable than
the first moment \cite{kme:fls}.  However, an attempt has been made to
include the theoretical uncertainty in the systematic errors as
described above.  The theoretical uncertainty from the lepton energy
moments is harder to estimate, because these moments are presently
calculated only to second order in $1/M_B$.  A resolution of the
discrepancy may require higher order expansions for the lepton energy
moments.  

Second, something may be wrong with the measurement of the lepton
energy moments.  Ligeti has questioned the lepton moment measurement
since it has some model dependence from the extrapolation below 0.6
GeV \cite{kme:ligeti}.  Another potential problem: there may be
additional sources of leptons in the CLEO data other than those
considered in \cite{kme:cleolep}.  Primary and secondary leptons are
separated using charge and kinematic correlations after removing
leptons from $\psi^{(')}$, $D_s$ and $\Lambda_c$ decays.  Besides
$\bar D$'s from the lower-vertex in $B$ decays (${\bar b}\to {\bar c}
W^+$), there can also be $D$'s from the upper-vertex ($W^+ \to c{\bar
s}$).  The contamination of the observable lepton spectrum from decays
of such upper-vertex $D$'s should be small because of the relatively
small branching fraction and the lower available energy given the
presence of two $D$ mesons in the decay.  However, in light of recent
results for upper-vertex $D$ production in $B$ decays
\cite{kme:uppervertex}, the contribution from this source of
background needs to be revisited.

Finally, it has been noted that moments of the photon energy spectrum
in $b\to s\gamma$ could provide constraints on $\bar\Lambda$ and
$\lambda_1$ \cite{kme:bauer}.  For example, the width of the photon
spectrum (measured in the $B$ rest frame) probes the $b$-quark motion,
{\em i.e.} $\lambda_1$.  But, it is important to note that the
backgrounds in $b\to s\gamma$ are very large, requiring an
experimental cut on the photon energy \cite{kme:jaffe}.

If the discrepancy between the two moment techniques remains after
further analysis, we may have to question the assumption of
quark-hadron duality implicit in such inclusive analyses.

\section{Conclusion}
The CLEO measurements of the $B\to D^{(*)}\ell\nu$ form factors
and $q^2$ distribution in $B\to \rho\ell\nu$ show progress in the
experimental understanding of the dynamics of heavy quark decay.  This
understanding, coupled with more theoretical work, should make
possible more precise determinations $|V_{cb}|$ and $|V_{ub}|$.
Likewise moment based analyses of inclusive semileptonic $B$ decays
seek to use data to constrain theory parameters and thereby reduce the
uncertainties in extracting $|V_{cb}|$ from the inclusive rate for
$b\to c\ell\nu$.

\end{document}